%% file: paper.tex
\documentclass[runningheads,a4paper,11pt]{llncs}

\usepackage{fullpage}
\usepackage{amsmath,amsfonts,amssymb,amscd}
\usepackage{times}
\usepackage{hyperref}

\usepackage{graphicx}
\usepackage{subfigure}
\usepackage{algorithmic}
\usepackage{algorithm}

\newcommand{\set}[1]{\left \{{#1} \right \}}
\newcommand{\rb}[1]{\left |{#1} \right \rangle}

\newcommand{\union}{\ensuremath{\bigcup}}
\newcommand{\intersect}{\ensuremath{\bigcap}}
\newcommand{\cond}{\;|\;}

\newcommand{\polylog}{\ensuremath{\mbox{polylog}}}
\newcommand{\zo}{\ensuremath{ \{ 0, 1 \} }}
\newcommand{\zon}{\ensuremath{ \{ 0, 1 \}^n }}
\newcommand{\eps}{\ensuremath{\varepsilon}}

\newcommand{\bigo}[1]{\ensuremath{{{O} \left( {#1} \right)}}}

\newcommand{\bigomega}[1]{\ensuremath{{\Omega \left( {#1} \right)}}}

\newcommand{\eqdef}{\ensuremath{:=}}

\newcommand{\N}{\ensuremath{\mathbb{N}}}

\newcommand{\remove}[1]{}

\usepackage{cite}

\usepackage[capitalise]{cleveref}
\Crefname{theorem}{Theorem}{Theorems}

\usepackage{color}

\title{Exact quantum query complexity of weight decision problems via Chebyshev polynomials}

\titlerunning{}

\author{Xiaoyu He  \and Xiaoming Sun \and Guang Yang \and Pei Yuan}
\authorrunning{Xiaoyu He  \and Xiaoming Sun \and Guang Yang \and Pei Yuan}

\usepackage{url}

\institute{Institute of Computing Technology, Chinese Academy of Sciences \\
University of Chinese Academy of Sciences \\
\path|{hexiaoyu18s,sunxiaoming,yangguang01,yuanpei}@ict.ac.cn|
}

\begin{document}

\maketitle
\thispagestyle{empty}
\begin{abstract}\normalsize
The weight decision problem, which requires to determine the Hamming weight of a given binary string, is a natural and important problem
with lots of applications in cryptanalysis, coding theory, fault-tolerant circuit design and so on.
In this work, we investigate the exact quantum query complexity of weight decision problems, where the \emph{exact} quantum algorithm must always output the correct answer within finite steps.
More specifically we consider a partial Boolean function which distinguishes the Hamming weight of the length-$n$ input
between $k$ and $l$.
In particular, both Deutsch-Jozsa problem
and Grover search problem can be interpreted as 
special cases of this problem. Our contributions include both upper bounds and lower bounds for the precise number of queries.
For most choices of $(k,l)$(or $(\frac{k}{n},\frac{l}{n})$) and sufficiently large $n$, the gap between our upper and lower bounds is no more than one query.
To get the results, we first build the connection between Chebyshev polynomials and our problem,
then determine all the boundary cases of $(\frac{k}{n},\frac{l}{n})$ with matching upper and lower bounds,
and finally we generalize to other cases via a new \emph{quantum padding} technique.
This quantum padding technique can be of independent interest in designing other quantum algorithms.
\end{abstract}


\section{Introduction}
\label{sec:intro}

\input{intro}

\section{Preliminaries}
\label{sec:pre}
\input{preliminary}

\section{Exact quantum algorithms for weight decision problems}
\label{sec:upp}
\input{upperbound}

\section{Exact quantum query complexity lower bounds}
\label{sec:low}
\input{exact_proof}

\section{Conclusion}
\label{sec:con}

In this work, we obtained both upper and lower bounds for the exact quantum query complexity of the weight decision function $f_n^{k,l}$.
For the lower bounds, we establish the relation between weight decision problems and Chebyshev polynomials, and combine it with the polynomial method (see Lemma \ref{beals}) to complete the proof.
That relation also leads to exact quantum algorithms with \emph{optimal} number of queries.
However, Chebyshev polynomials only provide bounds for $Q_E(f_n^{k,l})$ over very sparse choices of $n,k$, and $l$, among which most of ratios $\frac{k}{n}$ and $\frac{l}{n}$ are irrational and hence meaningless, even if allowing a traditional (naive) padding.
To that end we devise a ``quantum padding'' technique that is able to effectively pad an arbitrary number of zeros and ones and to make use of the quantum algorithms distinguishing non-integer weights.
With the new quantum padding technique, we develop a systematic method to determine upper and lower bounds for $Q_E(f_n^{k,l})$ by reducing to known bounds induced from Chebyshev polynomials. Furthermore, the upper and lower bounds are tight or nearly tight for a large fraction of choices of $(\frac{k}{n},\frac{l}{n})$.
Numerical calculation shows that the upper and lower bounds exactly match for more than $56\%$ area of $(\frac{k}{n},\frac{l}{n})$, and the gap is at most one for $97\%$ area
(as depicted in \cref{FIG}(b)).

One of the major future directions is to extend the knowledge of exact quantum query complexity to more complicated weight decision functions,
for example $f_n^{k,l,m}$ for distinguishing among three different weights or $f_n^{K,L}$ that distinguishes $|x|\in K$ from $|x|\in L$ for two disjoint subsets $K,L\subsetneq\set{0,1,\ldots,n}$.
It is also of interest to understand and fill up the gap between our upper and lower bounds and get a full characterization of $Q_E(f_n^{k,l})$,
which may require a deep understanding of the relation between exact quantum query complexity and the degree of corresponding (multi-linear/bounded forms/etc.) polynomials.

\appendix
\input{append}

\bibliographystyle{alpha}
\bibliography{exact_qa}

\end{document}

%% file: intro.tex


An important and amazing feature of quantum mechanics is that we can only ``read'' a quantum state by measuring it, when the superposed quantum state collapses to a \emph{random} sample according to a classical distribution.
As a result, many quantum algorithms would make error if the final quantum state does not collapse to the desired sample corresponding to the correct answer.
Although the error probability can be reduced through repetition, it remains natural to ask when and how such error can be completely eliminated.

\emph{Exact quantum algorithms} are quantum algorithms that always give exactly the correct answer (decoded from the measurement result). The complexity of such algorithms can be measured in the number of queries to the input oracle.
An example is the Deutsch-Jozsa algorithm \cite{deutsch1992rapid}
which shows an exponential speedup of quantum computation over classical ones.

\emph{Exact quantum query complexity} of a problem $\mathcal{P}$ is the necessary number of queries required by exact quantum algorithms to find the correct answer of $\mathcal{P}$.
This is a quantum analog of the classical decision tree complexity, and it is also the exact version of quantum query complexity where bounded-error are allowed.
Furthermore, the exact quantum query complexity turns out a good choice for demonstrating quantum advantage,
since its classical counterpart, the query complexity of deterministic algorithms, is usually significantly greater than that of (classical) bounded-error algorithms.
For example, a deterministic algorithm needs $\bigomega{n}$ queries to solve Deutsch-Jozsa problem whereas a trivial $\eps$-error \textrm{BPP} algorithm only spends $\bigo{\log\frac{1}{\eps}}$ queries.





The implementation of a quantum algorithm $\mathcal{A}$ in the query complexity model is described as follows:
$\mathcal{A}$ starts with a fixed state $\rb{\Psi_{start}}$ and performs a sequence of operations $U_0,O_x,U_1,O_x,\dots,O_x,U_t$, where every $O_x$ denotes an oracle query to the input $x$ and each $U_i$ is a unitary operator independent of $x$; the result $\mathcal{A}(x)$ is obtained from the measurement of the final state $\rb{\Psi_{end}} = U_t O_x U_{t-1}\cdots U_1 O_x U_0 \rb{\Psi_{start}}$.
The quantum query complexity of $\mathcal{A}$ is $t$ since it makes $t$ queries to the input oracle.
If furthermore $\mathcal{A}$ always give the correct answer,
$\mathcal{A}$ is an exact quantum algorithm with exact quantum query complexity $Q_E(\mathcal{A})=t$.
The exact quantum query complexity of a function $f$, denoted by $Q_E(f)$, is the minimum query complexity of all quantum algorithms that compute $f$ exactly on all inputs, {\it i.e.}
$$Q_E(f)=\min_{\mathcal{A}: \forall x, \mathcal{A}(x)=f(x)} Q_E(\mathcal{A})$$

In this paper we study the exact quantum query complexity of weight decision problems.
Such problems require to decide the Hamming weight $|x|$ of a vector $x$ drawn from $\zon$,
with all possible weights of $x$ given in advance.
In query complexity model this is equivalent to decide the output weight of a given Boolean function when interpreting $x$ as the truth table.
Such weight analysis of Boolean functions may find applications, as a whole or building blocks of more sophisticated algorithms, in various areas such as cryptanalysis \cite{filiol1998highly}, coding theory \cite{macwilliams1977theory}, fault-tolerant circuit design \cite{chakrabarty1996test}, the built-in self-testing of circuits \cite{chakrabarty1995cumulative} and so on.

In particular, two well-known quantum speedup examples are special cases of weight decision problems:
the Deutsch-Jozsa problem \cite{deutsch1992rapid} requires to distinguish $|x|=n/2$ from $|x|\in\set{0,n}$;
and the (decision version of) Grover search problem \cite{grover1996fast} distinguishes $|x|=0$ from $|x|=1$ (see \cite{brassard2002quantum,hoyer2000arbitrary,long2001grover} for exact algorithms).

Many previous studies on weight decision problems generalized the above two famous problems.
For example, Montanaro, Jozsa, and Mitchison \cite{montanaro2015on} considered the discrimination of $|x|=\frac{n}{2}$ from $|x|\in\set{0,1,n-1,n}$.
Qiu and Zheng \cite{qiu2016characterizations} proved that the exact quantum query complexity of distinguishing $|x|=\frac{n}{2}$ from $|x|\in\set{0,1,\dots,k, n-k,n-k+1,\dots,n}$ is $k+1$.
The Grover search problem can be generalized as the discrimination of two specific weights $|x|=k$ and $|x|=l$, for $(k,l)$ other than $(0,1)$.
Brassard \textit{et al.} \cite{brassard2002quantum} introduced quantum amplitude amplification as a general and optimal solution for the case $k=0$,
and with similar intuition Choi and Braunstein \cite{choi2011quantum,choi2012optimality} obtained asymptotically tight bounds for $l>k>0$.


The basis of our lower bound results is the polynomial method introduced by Beals \textit{et al.} \cite{beals2001quantum}. They prove that $Q_E(f)\ge \frac{1}{2}\deg(f)$ for every total Boolean function $f$. Here $\deg(f)$ denotes the degree of polynomial representation of $f$.
%
The tightness of such polynomial degree lower bounds has been studied but only known for bounded-error quantum algorithms so far.
Aaronson \textit{et al.} \cite{aaronson2015polynomials} considered the conversion from (bounded) polynomials to bounded-error quantum algorithms for partial Boolean functions and in particular they obtained an equivalence between $1$-query quantum algorithms and degree-$2$ polynomials.
However, a recent work by Arunachalam \textit{et al.} \cite{arunachalam2017quantum} showed that many degree-$4$ polynomials are far from functions computed by $2$-query quantum algorithms, \textit{i.e.} the equivalence between $k$-query quantum algorithms and degree-$2k$ polynomials does not hold in general.
This is evidence that lower bounds from polynomial degree may not be tight for quantum query complexity,
and partly explains the small gap between our upper and lower bounds.

\subsection{Main results}
In this paper we focus on the exact quantum query complexity of distinguishing two different weights.
We study the partial Boolean function $f_n^{k,l}:\zon\to\set{0,1}$ as below:
$$f_{n}^{k,l}(x)=\left\{
                     \begin{array}{ll}
                       0 & \mbox{if~} |x|=k; \\
                       1 & \mbox{if~} |x|=l; \\
                       undefined, & \hbox{otherwise.}
                     \end{array}
                   \right.
$$
In the definition of $f_n^{k,l}$ we put no further restrictions on $n,k$ and $l$, except that we assume they are integers satisfying $0\le k<l\le n$.
Our algorithm must give the correct answer of $f_n^{k,l}(x)$ after a fixed number of queries when $|x|\in\set{k,l}$, \textit{i.e.} no repetition or error reduction at all.
On the other hand, the output of our algorithm can be arbitrary for  $|x|\notin\set{k,l}$ since it is undefined in $f_n^{k,l}$.

\begin{figure}[!htb]
 \centering
 \centerline{\includegraphics[height=12cm,width=12cm]{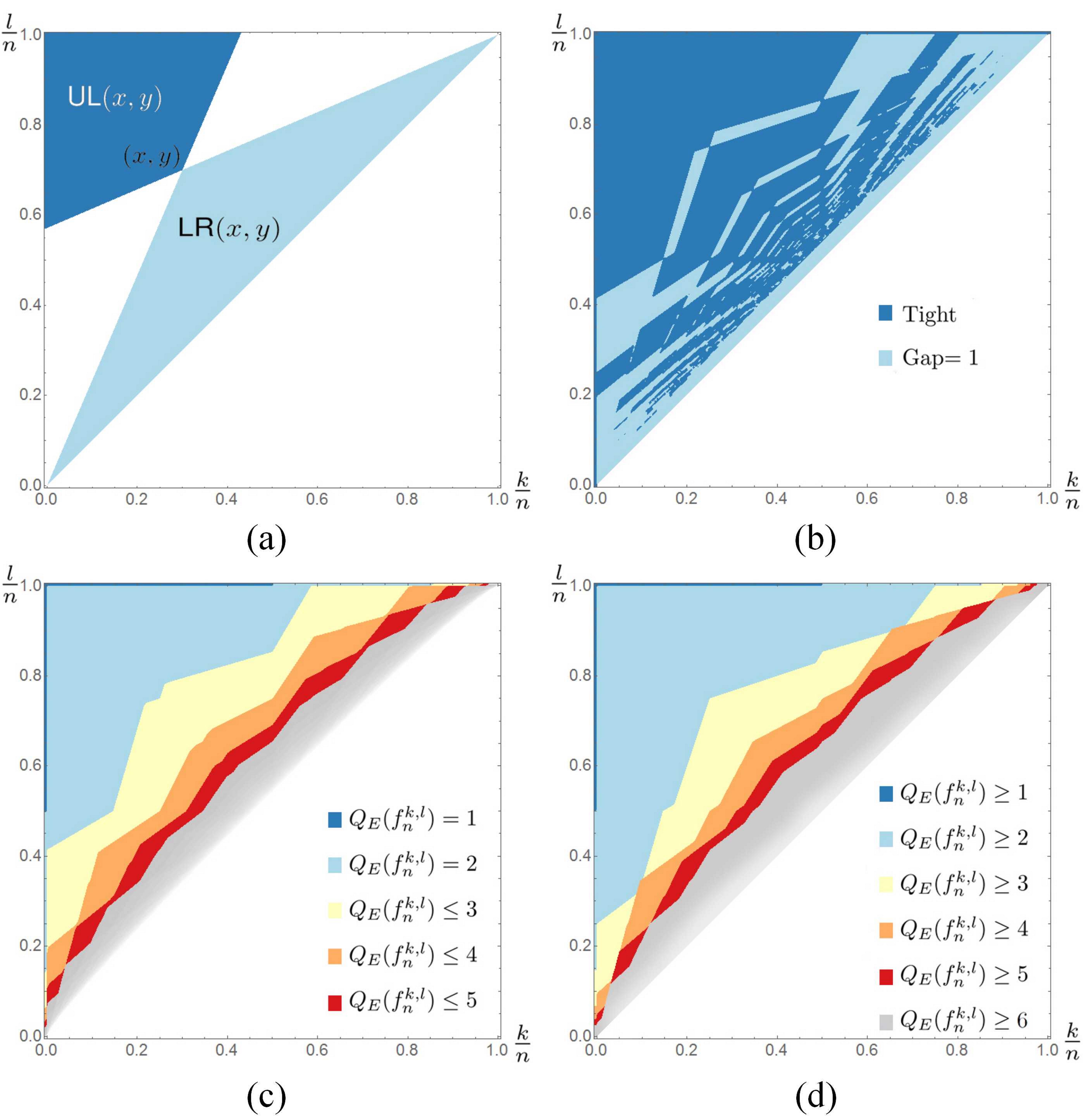}}
 \caption{(a) An example of upper-left region $\textsf{UL}(x,y)$ and lower-right region $\textsf{LR}(x,y)$ for $(x,y)=(0.3,0.7)$. $\textsf{UL}(x,y)$ includes the boundary of this region. $\textsf{LR}(x,y)$ includes the boundary expect the point $(x,y)$ and diagonal.
 (b) The gap between upper and lower bounds of $Q_E(f_n^{k,l})$ with respect to $\frac{k}{n}$ and $\frac{l}{n}$ and sufficiently large $n$.
 (c) The upper bounds for $Q_E(f_n^{k,l})$ with respect to $\frac{k}{n}$ and $\frac{l}{n}$, and sufficiently large $n$ (up to 5 queries).
 (d) The lower bounds for $Q_E(f_n^{k,l})$ with respect to $\frac{k}{n}$ and $\frac{l}{n}$ and sufficiently large $n$. (lower bounds up to $6$ queries)}
 \label{FIG}
\end{figure}

To characterize the effect of padding, we introduce the notion of \emph{upper-left region} and \emph{lower-right region} for every $(x,y)\in I^2$ where $x<y$ and $I\eqdef [0,1]$.
The upper-left region $\textsf{UL}(x,y)$ and the lower-right region $\textsf{LR}(x,y)$ are as follows (see \cref{FIG}(a) for a visual depiction):
$$                  \begin{array}{ll}
                        & \textsf{UL}(x,y)\eqdef \left\{(\kappa,\lambda)\in I^2|\lambda x\geq \kappa y ,(1-\kappa)(1-y)\geq(1-\lambda)(1-x),\kappa<\lambda\right\}; \\
                        & \textsf{LR}(x,y) \eqdef \left\{(\kappa,\lambda)\in I^2|\lambda x\leq \kappa y ,(1-\kappa)(1-y)\leq(1-\lambda)(1-x),\kappa<\lambda,(\kappa,\lambda)\neq (x,y)\right\}.
                     \end{array}
$$
Then we extend the definition of $\textsf{UL}$ and $\textsf{LR}$ to every set $S\subseteq I^2$ as $\textsf{UL}(S)\eqdef \bigcup_{(x,y)\in S}\textsf{UL}(x,y)$ and $\textsf{LR}(S)\eqdef \bigcup_{(x,y)\in S}\textsf{LR}(x,y)$.
Intuitively, for integers $0\le k<l\le n$ and $\kappa=\frac{k}{n}$, $\lambda=\frac{l}{n}$,
if $(\kappa,\lambda)\in \textsf{UL}(x,y)$, then any exact quantum algorithm which solves $f_{n'}^{k',l'}$ for $(\frac{k'}{n'},\frac{l'}{n'})=(x,y)$ will induce an exact quantum algorithm solving $f_n^{k,l}$ after padding some zeros and ones to the input of $f_n^{k,l}$. Therefore, $Q_E(f_n^{k,l})\leq Q_E(f_{n'}^{k',l'})$.
Similar reduction holds for $(\kappa,\lambda)\in \textsf{LR}(x,y)$,
when $(x,y)\in \textsf{UL}(\kappa,\lambda)$.

Now we introduce the definition of $S_d$ consisting of the boundary cases which we can solve with $d$-query exact quantum algorithms.
Indeed every element in $S_d$ corresponds to a pair of consecutive extrema of degree-$D$ Chebyshev polynomial,
where $D=2d$ or $D=2d-1$ (more details of $S_d$ are discussed in \cref{sec:pre}).
For every $d\in\mathbb{N}$, we define $S_d$ as below:
\begin{multline}\label{def:Sd}
S_d  \eqdef \set{\left( \frac{1}{2}{\left(1-\cos\left(\frac{\gamma\pi}{2d}\right)\right)},\frac{1}{2}{\left(1-\cos\left(\frac{(\gamma+1)\pi}{2d}\right)\right)}   \right) \;\Big|\; \gamma\in\set{0,1,\dots,2d-1}} \\
  \bigcup \set{\left( \frac{1}{2}{\left(1-\cos\left(\frac{\gamma\pi}{2d-1}\right)\right)},\frac{1}{2}{\left(1-\cos\left(\frac{(\gamma+1)\pi}{2d-1}\right)\right)}   \right) \;\Big|\; \gamma\in\set{1,\dots,2d-3}}.
\end{multline}

Our first main contribution is the construction of a family of exact quantum algorithms which immediately implies the following theorem.

\begin{theorem}[Upper bounds]\label{padding}
For every $d\in\mathbb{N}$ and $0\leq k<l\leq n$ with $k,l,n\in\mathbb{N}$, let $\kappa=\frac{k}{n}$ and $\lambda=\frac{l}{n}$. If $(\kappa,\lambda)\in \mathsf{UL}(S_d)$, then $Q_E(f_n^{k,l})\leq d.$
\end{theorem}
Roughly speaking, the upper bound of $Q_E(f_n^{k,l})$ is determined by all elements of $S_d$ that $f_n^{k,l}$ can be reduced to via an enhanced ``quantum padding'' technique, since every case in $S_d$ can be solved exactly with $d$ quantum queries.

Unlike the na\"{i}ve padding where the number of padded zeros and ones must be non-negative integers,
our quantum padding technique can effectively pad an arbitrary (even irrational) non-negative number of zeros and ones.
As a result, $Q_E(f_n^{k,l})$ has an upper bound fully and smoothly determined by $\frac{k}{n}$ and $\frac{l}{n}$.
However, the non-negativity requirement remains and restricts the power of our padding technique to the upper-left region, {\it e.g.}  $\textsf{UL}(S_d)$.

The upper bound for $Q_E(f_n^{k,l})$ in \cref{padding} can be represented in terms of $k$ and $l$ as in \cref{asym}.
This is asymptotically optimal since there is a matching lower bound for (bounded-error) quantum query complexity of $f_n^{k,l}$ by \cite{choi2012optimality}.

\begin{corollary}\label{asym}
If $k,l,n\in\mathbb{N}$ and $0\le k< l\le n$, then $Q_E(f_n^{k,l})=O\left(\frac{\sqrt{(n-k)l}}{l-k}\right)$.
\end{corollary}


Moreover, \cref{padding} implies quantum advantage of weight decision problems in the communication complexity model.
In the context of communication complexity \cite{yao1979some},
two parties Alice and Bob with input $a\in A$ and $b\in B$ respectively want to compute a function $F:A\times B\to \zo$ with minimum communication, \textit{i.e.} only the number of exchanged bits matters.
In the quantum communication complexity \cite{brassard2003quantum,buhrman2010nonlocality,hromkovivc2013communication,kushilevitz1997communication}, Alice and Bob are allowed to exchange quantum states stored in qubits,
and hence it becomes possible to save communication.
\cref{corollary:CC}, which follows \cref{padding}, gives an $O(d\log n)$ upper bound for the quantum communication of a two-party weight decision function $F_n^{k,l}(a,b)\eqdef f_n^{k,l}(a\oplus b)$,
where $A=B=\zon$ and $a\oplus b$ refers to the bitwise parity.
Note that for small $d$ this is an exponential speedup on
the (classical) deterministic communication complexity of $F_n^{k,l}$ (which is $n-l+k+1$).
This also improves on the previous result by Gruska \emph{et al.} \cite{gruska2017generalizations}, which requires $O(\log n)$ quantum communication to compute $F_n^{0,k}$ when $k\geq \frac{n}{2}$.


\begin{corollary}\label{corollary:CC}
 For every $d\in\mathbb{N}$ and $0\leq k<l\leq n$ with $k,l,n\in\mathbb{N}$, let $\kappa=\frac{k}{n}$ and $\lambda=\frac{l}{n}$. If $(\kappa,\lambda)\in \mathsf{UL}(S_d)$, then the exact quantum communication complexity of $F_n^{k,l}$ is $O(d \log n )$.
\end{corollary}

On the lower bound part, we prove \cref{lowerbound} with polynomial method and padding.
The polynomial method \cite{beals2001quantum} relates quantum query complexity to the degree of polynomial representations.
For the weight decision function $f_n^{k,l}$, we discover its relation to extrema of Chebyshev polynomials,
and prove the exact quantum query lower bound for elements in $S_d$ via a degree analysis.
Finally we apply the same padding technique as before (but in the other direction) for generalization.
The family of sets $\set{S_d}_{d\in\N}$ is exactly the same as defined in \cref{def:Sd} and used in \cref{padding}.
\begin{theorem}[Lower bounds]\label{lowerbound}
For every $d\in\mathbb{N}$ and $0\leq k<l\leq n$ with $k,l,n\in \mathbb{N}$, let $\kappa=\frac{k}{n}$ and $\lambda=\frac{l}{n}$. If $(\kappa,\lambda)\in \mathsf{LR}(S_d)$, then $Q_E(f_n^{k,l})\geq d+1$ for sufficiently large $n$.
\end{theorem}

Note that the lower bound of $Q_E(f_{n}^{k,l})$ is fully determined by $\kappa=\frac{k}{n}$ and $\lambda=\frac{l}{n}$ when $n$ is sufficiently large. A visual depiction of its variation over $(\kappa,\lambda)$ for $0\le \kappa<\lambda\le 1$ is given in \cref{FIG}(d).
Combining \cref{padding} and \cref{lowerbound}, we get both upper and lower bounds for $Q_E(f_n^{k,l})$ by finding the corresponding $S_{d_1}$ and $S_{d_2}$ such that $d_1+1\le Q_E(f_n^{k,l})\le d_2$,
\textit{cf.} \cref{FIG}(c) and \cref{FIG}(d).
By a numerical calculation,
we find that our upper bounds and lower bounds are nearly optimal -- the bounds exactly match for $> 56\%$ area in the figures, and the gap is no more than one for $> 97\%$ area, as depicted in \cref{FIG}(b).

Furthermore, our proofs of \cref{padding} and \cref{lowerbound}
also work for the partial Boolean $g_n^k$,
where $g_n^k(x)=0$ for $|x|\in\{k,n-k\}$ and $g_n^k(x)=1$ for $|x|=\frac{n}{2}$.
In particular, $Q_E(g_n^k)=Q_E(f_n^{k,n/2})=Q_E(f_n^{n/2,n-k})$.
See \cref{specialfunction} for more details.

\subsection{Comparison to related works}
%
The weight decision function $f_n^{k,l}$ studied in this paper is most relevant to the generalization of  Grover search \cite{grover1996fast} and amplitude amplification \cite{brassard2002quantum} which distinguishes two specific weights $k$ and $l$.
Along this line, Braunstein \textit{et al.} \cite{braunstein2007exact} first proved that
$Q_E(f_n^{k,n-k})=O(\frac{n}{|n-2k|})$.
Later, Choi and Braunstein \cite{choi2011quantum} showed $Q_E(f_n^{k,l})=O(\frac{n}{l-k})$ when $0\le k<\frac{n}{2}<l\le n$ via a reduction to the symmetric form $f_n^{k,n-k}$,
and Choi \cite{choi2012optimality} proved the asymptotic optimality of this upper bound (for $0\le k<\frac{n}{2}<l\le n$) with techniques from \cite{ambainis2000quantum}.
Qiu and Zheng \cite{qiu2016characterizations} proved that $Q_E(f_n^{\frac{n}{4},\frac{3n}{4}})=2$ and characterized all cases reducible to $f_n^{\frac{n}{4},\frac{3n}{4}}$ via classical padding.
This does save one query compared with \cite{choi2011quantum}, but it turns out just a special case of our algorithm and  further requires $l-k$ to be even to apply a classical padding.
\cite{qiu2016characterizations} also proved $Q_E(f_n^{0,k})=2$ for $\frac{n}{4}\le k<\frac{n}{2}$, extending the previous knowledge \cite{gruska2017generalizations} that $Q_E(f_n^{0,l})=1$ for $l\ge \frac{n}{2}$.
A very recent work by Scott Aaronson \cite{aaronson2018quantum} studied the tradeoff between two kinds of quantum resources for computing $f_n^{k,2k}$ with bounded error, and proved the necessity of either $\Omega(\sqrt{n/k})$ queries or $\Omega(\min\{k^{1/4},\sqrt{n/k}\})$ copies of $|x\rangle=\frac{1}{\sqrt{|x|}}\sum_{x_i=1}|i\rangle$, which is technically orthogonal to our work.
In summary, previous results on quantum query complexity of $f_n^{k,l}$ are either asymptotic or ad-hoc, and the crude padding technique only handles discrete parameters and sometimes requires divisibility.
In this paper we not only propose a systemic solution that vastly extends the knowledge of (both upper and lower bounds of) $Q_E(f_n^{k,l})$,
but also introduce a quantum padding technique that removes range and divisibility restrictions such as $k<\frac{n}{2}<l$ or $l-k$ must be even.

We remark that the bounded-error quantum query complexity of $f_n^{k,l}$ is  $Q_2(f_{n}^{k,l})=\Omega\left(\frac{\sqrt{(n-k)l}}{l-k}\right)$ by \cite{choi2012optimality}, which matches our upper bound for $Q_E(f_n^{k,l})$ as in \cref{asym}.
Thus both exact and bounded-error quantum query complexity of $f_n^{k,l}$ are $\Theta\left(\frac{\sqrt{(n-k)l}}{l-k}\right)$,
with an exponential advantage over the deterministic query complexity $D(f_{n}^{k,l})=n-l+k+1$.

Total Boolean functions of similar forms are also studied but in a way technically orthogonal to our work.
For example, Ambainis \textit{et al.} \cite{ambainis2016exact} proved that $\max\set{n-k,l}- 1\le Q_E(\mathsf{EXACT}_{k,l}^n)\le \max\set{n-k,l}+ 1$,
where $0\le k\le l\le n$, $\mathsf{EXACT}_{k,l}^n(x)=1$ on inputs of Hamming weight $|x|\in\set{k,l}$ and $\mathsf{EXACT}_{k,l}^n(x)=0$ otherwise.
The quantum advantage of $\mathsf{EXACT}_{k,l}^n$ is much less than that of $f_n^{k,l}$ mainly because it is a total Boolean function.
In general, the quantum speedup for total Boolean functions are polynomially bounded:
in the bounded-error setting, Beals \textit{et al.} \cite{beals2001quantum} proved that $Q_2(f)=\bigomega{D(f)^{1/6}}$; 
and Midrijanis \cite{midrijanis2004exact} proved that $Q_E(f)=\bigomega{D(f)^{1/3}}$.
However, the best separation from randomized query complexity was linear until 2013 when the first superlinear speedup example with $Q_E(f)=\bigo{R_2(f)^{0.86\dots}}$ was discovered by Ambainis \cite{ambainis2013superlinear} (the current best separation is $Q_E(f)=\widetilde{O}\left(R_2(f)^{1/2}\right)$ by \cite{ambainis2016separations}).
In \cite{ambainis2014exact}, Ambainis \textit{et al.} showed that almost all $n$-bit Boolean functions have exact quantum query complexity less than $n$ expect $\mathsf{AND}_n$, up to isomorphism.
Aaronson \textit{et al.} \cite{aaronson2016separations} presented a total function $f$ with $R_2(f)=\widetilde{\Omega}(Q_2(f)^{2.5})$.
Aaronson and Ambainis \cite{aaronson2009need} together proved that $R_2(f)=O(Q_2(f)^7\polylog Q_2(f))$ for every symmetrically partial  (not necessarily Boolean) function $f$.

\subsection{Organization}
The rest part of this paper is organized as follows.
In \cref{sec:pre} we introduce necessary notations and definitions.
In \cref{sec:upp} we present our constructions of exact quantum algorithms for the weight decision problems and prove upper bounds.
Our lower bound results are exhibited in \cref{sec:low}.
Finally we summarize the paper in \cref{sec:con}.

%% file: preliminary.tex
For any positive integer $t$, let $[t]\eqdef\set{1,\ldots,t}$ and $[t]_0 \eqdef \set{0,1,\ldots,t}$.
Let $f: D \rightarrow \zo$ denote a (partial) Boolean function whose domain is $D\subseteq \zon$. In the quantum query model,
$O_x$ is the quantum oracle query to the input $x$ defined as
$$O_x|i\rangle \eqdef \left\{
                     \begin{array}{ll}
                       (-1)^{x_i}|i\rangle,  & \text{if } i\in[n];\\
                       |0\rangle,  & \text{if } i=0.
                     \end{array}
                   \right.
$$
We say that a quantum algorithm computes $f$ exactly if for every $x\in D$ the algorithm outputs $f(x)$ with probability $1$ in finite steps.
The exact quantum query complexity $Q_E(f)$ is the minimum number of queries of all quantum algorithms computing $f$ exactly.

It is easy to show that every Boolean function can be represented by $n$-variate polynomials.
\begin{definition}\label{degree}
For a function $f:D \rightarrow \zo $ with domain $D\subseteq \zon$,
an $n$-variate polynomial $p:\mathbb{R}^n\rightarrow \mathbb{R}$ is called a \emph{real-valued multilinear polynomial representation} of $f$ if $f(x)=p(x)$ for all $x\in D$ and $0\le p(x) \le 1$ for all $x\in \zon$.
The \emph{degree} of $f$, denoted by $\deg(f)$, is defined as the minimum degree of all real-valued multilinear polynomial representations of $f$.
\end{definition}

\vspace{-2mm}
A Boolean function is \emph{symmetric} if permutating its input does not change its output. Namely, the value of $f(x)$ is fully determined by $|x|$.
In fact, any real-valued multilinear polynomial representation of a symmetric function can be converted to a univariate polynomial of $|x|$ with the same degree \cite{minsky1988perceptrons}. We will make no distinction between these two polynomial representation of symmetric Boolean functions if the meaning is clear from the context.
%
%

The following lemma by Beals \textit{et al.} establishes a connection between
the degree of $f$ and its exact quantum query complexity $Q_E(f)$.

\begin{lemma}[\cite{beals2001quantum}]\label{beals}
If $f$ is a (total) Boolean function, then $Q_E(f)\ge \deg(f)/2$.
\end{lemma}

According to the following Property \ref{symmetric}, we will only investigate $f_n^{k,l}$ with $k<l$.
\begin{property}[\cite{qiu2016characterizations}]\label{symmetric}
For every $n\in \mathbb{N}$ and $k,l\in [n]_0$, $Q_E(f_n^{k,l})= Q_E(f_n^{n-l,n-k})$.
\end{property}

Suppose that $f_{n'}^{\kappa n',\lambda n'}$ is computed by a $d$-query exact quantum algorithm. To compute $f_n^{k,l}$, we can pad $a$ many zeros and $b$ many ones to the input of $f_n^{k,l}$ and hence reduce $f_n^{k,l}$ to $f_{n+a+b}^{k+b,l+b}$, where the latter is indeed $f_{n'}^{\kappa n',\lambda n'}$ when $a=\frac{l-k}{\lambda-\kappa}-\frac{l\kappa-k\lambda}{\lambda-\kappa}-n$ and $b=\frac{l\kappa-k\lambda}{\lambda-\kappa}$ (so that $k+b=\kappa n'$ and $l+b=\lambda n'$ for $n'=n+a+b$).
As long as $a,b,k$ and $l$ are non-negative integers, 
the above reduction is straightforward and $f_n^{k,l}$ can be computed by a $d$-query exact quantum algorithm as well, \textit{i.e.} $Q_E(f_n^{k,l})\le Q_E(f_n^{\kappa n.\lambda n})$.
However the na\"{i}ve padding technique is very limited in the sense that we cannot pad a non-integer (or irrational, if we only care about $\frac{k}{n}$ and $\frac{l}{n}$) number of zeros or ones.
The quantum implementation of ``padding'' is discussed in details in \cref{general}.

\vspace{1mm}
\noindent\textbf{Chebyshev polynomials. }
 In this paper, we employ the mathematical tools {\em Chebyshev polynomials of the first kind} \cite{askey1991chebyshev}, which are denoted by $\{T_n\}_{n=1}^\infty$.
 More specifically, we use the following properties of the Chebyshev polynomials:
\begin{property}[\cite{askey1991chebyshev}]\label{chebyproperty}
Let $T_n$ be the first kind Chebyshev polynomial of degree $n$, then
\begin{enumerate}
  \item  $|T_n(x)|\leq 1$ for $|x|\le 1$, and $|T_n(x)|>1$ for $|x|>1$;
  \item  $T_n(\cos(\frac{k\pi}{n}))=(-1)^k$ for  $k\in[n]_0$ and  $T_n'(\cos(\frac{k\pi}{n}))=0$ for $k\in[n-1]$.
\end{enumerate}
\end{property}


For convenience, we call the points in $\set{x\in[-1,1]\cond |T_n(x)|=1 }$ \emph{extrema} of $T_n$.

Now we explain the relation between Chebyshev polynomials and the weight decision problem.
Noticing that the range of Chebyshev polynomials is $[-1,1]$, we consider another representation of Boolean functions on $\set{-1,1}$, \emph{i.e.}
$\hat{f}:\set{-1,1}^n\to\set{-1,1}$.
The function $f_n^{k,l}:\zon\to\set{0,1}$ can be transformed into $\hat{f}_n^{k,l}$ defined on input $\hat{x}=(\hat{x}_1,\ldots,\hat{x}_n)\in\{-1,1\}^n$ by setting $\hat{x}_i=(-1)^{x_i}$ for every $x_i\in\zo$ and $i\in[n]$:
$$\hat{f}_{n}^{k,l}(\hat{x})=\left\{
                     \begin{array}{ll}
                       1, & \mbox{if~} |\hat{x}|=n-2k; \\
                      -1, & \mbox{if~} |\hat{x}|=n-2l;\\
                      undefined, & \hbox{otherwise.}
                     \end{array}
                   \right.
$$
Note that $\hat{f}_n^{k,l}$ distinguishes $|\hat{x}|=n-2k$ from $|\hat{x}|=n-2l$ since $|\hat{x}|\eqdef\sum_{i=1}^n\hat{x}_i=n-2|x|$.

The bijection between $f_n^{k,l}(x)$ and $\hat{f}_n^{k,l}(\hat{x})$ also holds on their polynomial representations, \emph{i.e.} if a polynomial $p(|x|)$ represents $f_n^{k,l}(x)$ then  $\hat{p}(|\hat{x}|)=1-2\cdot p\left(\frac{n-|\hat{x}|}{2}\right)$ represents $\hat{f}_n^{k,l}(\hat{x})$ accordingly.
And in particular, $\hat{p}(|\hat{x}|)=\hat{f}_n^{k,l}(\hat{x})$ for $|\hat{x}|=n-2k$ and $|\hat{x}|=n-2l$, and $-1\leq \hat{p}(|\hat{x}|)\leq 1$ for every $\hat{x}\in\{-1,1\}^n$. 
However, the domain of $\hat{p}$ is $[-n,n]$ since $|\hat{x}|\le n$ for $\hat{x}\in\{-1,1\}^n$.
To fit the Chebyshev polynomials which are bounded on the domain $[-1,1]$, we introduce the polynomial $\hat{P}$ such that $\hat{P}(z)=\hat{p}(z n)$.
Note that $\deg(\hat{P})=\deg(\hat{p})=\deg(p)$ since we only use linear transformations.
In particular $\hat{P}(1-\frac{2k}{n})=\hat{p}(n-2k)=1$ corresponds to $p(k)=0$, $\hat{P}(1-\frac{2l}{n})=\hat{p}(n-2l)=-1$  corresponds to $p(l)=1$.

Let $\hat{P}$ be the Chebyshev polynomial of degree $D=2d$ or $2d-1$, \emph{i.e.} $T_D=T_{2d}$ or $T_{2d-1}$.
Then $\hat{P}$ should distinguish the two weights as $\hat{P}(1-\frac{2k}{n})=(-1)^\gamma$ and  $\hat{P}(1-\frac{2l}{n})=(-1)^{\gamma+1}$, which turn out two extrema of the Chebyshev polynomial $T_D$.
Recalling that padding is inherently limited in amplifying gaps, we consider consecutive extrema of $T_D$ in the form $(\eta_\gamma,\eta_{\gamma+1})\eqdef \left(\cos(\frac{\gamma\pi}{D}),\cos(\frac{(\gamma+1)\pi}{D})\right)$ as boundary cases where $T_D(\eta_\gamma)=-T_D(\eta_{\gamma+1})$ and $T'_D(\eta_\gamma)=T'_D(\eta_{\gamma+1})=0$.
Translating $(1-\frac{2k}{n},1-\frac{2l}{n})\equiv(\eta_\gamma,\eta_{\gamma+1})$ back to the polynomial $p$ representing $f_n^{k,l}$, we get $\left(\frac{k}{n},\frac{l}{n}\right)=\left( \frac{1}{2}(1-\cos(\frac{\gamma\pi}{D})), \frac{1}{2}(1-\cos(\frac{(\gamma+1)\pi}{D}))  \right)$ as a boundary case for the weight decision problem.
The definition of $\set{S_d}_{d\in\N}$ in \cref{def:Sd} consists of such boundary cases derived from extrema of both $T_{2d}$ and $T_{2d-1}$ for every $d$, after tailored\footnote{For $\gamma=0$ the case $\left(0,\frac{1}{2}\left(1-\cos\left(\frac{\pi}{2d-1}\right)\right)\right)$ is excluded since it is dominated by $\left(0,\frac{1}{2}\left(1-\cos\left(\frac{\pi}{2d}\right)\right)\right)$ and does not affect the bounds. The symmetric case $\left(\frac{1}{2}\left(1-\cos\left(\frac{(2d-2)\pi}{2d-1}\right)\right),1\right)$ is excluded as well.} for our upper and lower bound results.

%% file: upperbound.tex


In this section, we introduce \cref{padding} for the upper bound of the weight decision function $f_n^{k,l}$, and then prove it by constructing an exact quantum algorithm for $f_n^{k,l}$.

\noindent
 \textbf{\cref{padding}.}
\textit{For every $d\in\mathbb{N}$ and $0\leq k<l\leq n$ with $k,l,n\in\mathbb{N}$, let $\kappa=\frac{k}{n}$ and $\lambda=\frac{l}{n}$. If $(\kappa,\lambda)\in \mathsf{UL}(S_d)$, then $Q_E(f_n^{k,l})\leq d.$
}

Since $f_n^{k,l}$ is trivialized if any of $n,k,l$ is non-integer, we assume $n,k,l\in\N$ in the rest of this section.
Obviously the upper bound of $Q_E(f_n^{k,l})\le d$ is determined by the minimum $d$ satisfying the conditions in  \cref{padding}.
That is,
for $n,k,l$ satisfying $0\leq k<l\leq n$,
$Q_E(f_n^{k,l})\le \min\set{d \;|\; (\frac{k}{n},\frac{l}{n})\in \mathsf{UL}(S_d) }.$

\cref{padding} can be rewritten by representing $d$ as a function of $k$ and $l$,
and thus the upper bound of $Q_E(f_n^{k,l})$ becomes \cref{asym}.
This bound is asymptotically tight since it matches the lower bounds of bounded-error quantum query complexity of $f_n^{k,l}$.

\noindent
\textbf{\cref{asym}.}
\textit{If $k,l,n\in\mathbb{N}$ and $0\le k< l\le n$, then $Q_E(f_n^{k,l})=O\left(\frac{\sqrt{(n-k)l}}{l-k}\right)$.}

\begin{proof}
Let $\textsf{R}(x,y)=\left\{(\kappa,\lambda)\in I^2 | \kappa\leq x\wedge\lambda\geq y\right\}$ be the upper-left rectangle of $(x,y)$.
It immediately follows that $\textsf{R}(x,y)\subseteq \textsf{UL}(x,y)$.
By \cref{padding}, if $(x,y)\in S_d$ and $(\frac{k}{n},\frac{l}{n})\in \textsf{R}(x,y)\subseteq \textsf{UL}(x,y)$, then there exists a $d$-query exact quantum algorithm for $f_n^{k,l}$, {\it i.e.} $Q_E(f_n^{k,l})\le d$.
In what follows we will determine an upper bound for $d$.

In order to get $(\frac{k}{n},\frac{l}{n})\in \textsf{R}(x,y)$ for some $d$ and $(x,y)\in S_d$,
it suffices to find  $\gamma\in [2d-1]_0$ such that $x=\frac{1}{2}\left(1-\cos \Big(\frac{\gamma \pi}{2d}\right)\Big)$ and $y=\frac{1}{2}\left(1-\cos \left(\frac{(\gamma+1) \pi}{2d}\right)\right)$ satisfy the following:
$$\left\{\begin{array}{ll}\frac{k}{n}\leq x=\frac{1}{2}\left(1-\cos \Big(\frac{\gamma \pi}{2d}\right)\Big) =\sin^2 \left(\frac{\gamma\pi}{4d}\right); \\
\frac{l}{n}\geq  y =\frac{1}{2}\left(1-\cos \left(\frac{(\gamma+1) \pi}{2d}\right)\right) =\sin^2 \left(\frac{(\gamma+1)\pi}{4d}\right).
\end{array}\right.$$
Solving above inequalities for $\gamma$,
we get 
$$0\le \frac{4d}{\pi}\arcsin \sqrt{\frac{k}{n}}\leq \gamma\leq \frac{4d}{\pi}\arcsin\sqrt{\frac{l}{n}}-1\le 2d-1.$$

Therefore, the desired integer $\gamma\in [2d-1]_0$ exists if $d$ is sufficiently large such that
$$\left(\frac{4d}{\pi}\arcsin\sqrt{\frac{l}{n}} - 1\right)-\frac{4d}{\pi}\arcsin \sqrt{\frac{k}{n}}\geq 1,$$
which is
$d\ge \frac{\pi}{2} \Big/ \left(\arcsin\sqrt{\frac{l}{n}} - \arcsin \sqrt{\frac{k}{n}}\right).$
Next we show that $d = O\left(\frac{\sqrt{(n-k)l}}{l-k}\right)$ suffices if $k<\frac{n}{2}$.

Notice that $g(z)=\arcsin (z) -z$ is monotonically increasing when $z\in [0,1]$.
Thus we have $\arcsin \sqrt{\frac{l}{n}}-\arcsin\sqrt{\frac{k}{n}} > \sqrt{\frac{l}{n}}-\sqrt{\frac{k}{n}}>0$ as long as $0\le k<l\le n$, and hence
$$  \frac{\pi}{2} \Big/ \left(\arcsin\sqrt{\frac{l}{n}} - \arcsin \sqrt{\frac{k}{n}}\right)  < \frac{\pi}{2} \Big/ \left(\sqrt{\frac{l}{n}} -  \sqrt{\frac{k}{n}}\right) = O\left(\frac{\sqrt{n} (\sqrt{l}+\sqrt{k})}{l-k}\right)=O\left(\frac{\sqrt{(n-k)l}}{l-k}\right)$$
where the last equality holds because $\sqrt{l}+\sqrt{k} < 2\sqrt{l}$ for $k<l$
and $\sqrt{n}<2\sqrt{n-k}$ for $k<n/2$.

Therefore, for any $k<\frac{n}{2}$, there is $Q_E(f_n^{k,l})\le d=O\left(\frac{\sqrt{(n-k)l}}{l-k}\right)$.
Exactly the same upper bound holds for $k\ge \frac{n}{2}$ since $Q_E(f_n^{k,l})=Q_E(f_n^{n-l,n-k})$ and now $n-l<n-k\le \frac{n}{2}$.\qed
\end{proof}

To design an exact quantum algorithm for $f_n^{k,l}$, we recall the geometric view  (see \cite{grover1996fast,hoyer2000arbitrary,long2001grover,brassard2002quantum}) of quantum computation that unitary operators can be realized as rotations and only orthogonal subspaces can be perfectly distinguished.
In order to compute $f_n^{k,l}$, the initial state on inputs of Hamming weights $k$ and $l$ must be rotated to final states in orthogonal subspaces so that we can perfectly distinguish them.
In \cref{relationship}, we show how this can be done for $f_n^{k,l}$ with $\le d$ queries when $1-2\kappa$ and $1-2\lambda$ are consecutive extrema of a Chebyshev polynomial $T_{D}(x)$ for $\kappa =\frac{k}{n}$, $\lambda =\frac{l}{n}$ and $D\le 2d$.
More choices of $n,k,l$ can be handled by padding to known cases.
However, classical padding does not work in general since most extrema are  irrational and hence cannot be padded to classically.
In \cref{general}, we complete the proof of \cref{padding} with our ``quantum padding'' technique.
For accessibility we also give the pseudo-code of our algorithm in \cref{alg}.

\subsection{Extrema of Chebyshev polynomials and exact quantum algorithms}\label{relationship}
In this section, we will discuss the relationship between extrema of Chebyshev polynomials and exact quantum algorithms for $f_n^{k,l}$. In order to distinguish inputs with two different weights, we need a quantum algorithm that the final states are in two orthogonal subspaces corresponding to inputs of different weights.
Let $\kappa = \frac{k}{n}$ and $\lambda =\frac{l}{n}$ for $n,k,l\in\mathbb{N}$. Suppose $\delta\in\{0,1\}$, $d\in[n]$ and $\gamma,\chi\in[2d-\delta-1]_0$.
In what follows, we will show that if $1-2\kappa$ and $1-2\lambda$ are two extrema $\eta_\gamma\eqdef \cos\left(\frac{\gamma\pi}{2d-\delta}\right),\eta_{\chi}\eqdef \cos\left(\frac{\chi\pi}{2d-\delta}\right)$  of a Chebyshev polynomial $T_{2d-\delta}$ and $\gamma-\chi$ is odd, then $\set{x\in\zon\;|\; |x|=k}$ and $\set{x\in\zon\;|\; |x|=l}$ can be distinguished by a $d$-queries quantum algorithm.
For simplicity we only consider the case $\gamma-\chi=1$ in this subsection,
and defer the discussion of the case $\gamma-\chi>1$ to \cref{general} where the exact quantum algorithm may use no more than $d$ queries by quantum padding technique.

Let the initial state be $|\Psi_0\rangle = \frac{1}{\sqrt{n}}\sum_{i=0}^n \rb{i}$ and $\theta=\arcsin \sqrt{\frac{|x|}{n}}$, then $\rb{\Psi_0}$ can be interpreted as follows:
$$|\Psi_0\rangle=\cos\theta|\alpha_\perp\rangle+\sin\theta|\alpha\rangle$$
where $|\alpha_\perp\rangle\eqdef\frac{1}{\sqrt{n-|x|}}\sum_{i:x_i=0}|i\rangle$ and $|\alpha\rangle\eqdef\frac{1}{\sqrt{|x|}}\sum_{i:x_i=1}|i\rangle$.
For the construction of our algorithm we define two unitary transformations $W$ and $U$ as follows.
\begin{itemize}
  \item [(1)] $W$ is a unitary transformation over a $n$-dimensional Hilbert space with basis vectors $\{|1\rangle,\ldots,|n\rangle\}$. It is a unitary transformation described as follows:
$$W|k\rangle=\frac{2}{n}\sum_{i=1}^n|i\rangle-|k\rangle,~\forall k\in[n].$$
  \item [(2)] $U$ denotes a unitary transformation over a $\left(n+\binom{n}{2}\right)$-dimensional Hilbert space with basis vectors $\{|k\rangle,|i,j\rangle \mid i,j,k \in[n],i<j\}$. It is a unitary completion of the following  transformation:
$$U|k\rangle=\frac{1}{n}\sum_{i=1}^n|i\rangle+\frac{1}{\sqrt{n}}\left(\sum_{i:k<i\leq n}|k,i\rangle-\sum_{i:1\leq i <k}|i,k\rangle\right),~\forall k\in[n].$$
\end{itemize}

It is easy to verify that both $W$ and $U$ are unitary transformations. For convenience, we further define two unitary transformations $G$ and $R$ such that $G\eqdef WO_x$ and $R\eqdef UO_x$. The operator $G$ is also known as the Grover operator \cite{grover1996fast} and $R$ is a rotation operator. After applying the operators $G$ and $R$ respectively, the initial state $\rb{\Psi_0}$ becomes
$$G|\Psi_0\rangle=\cos(3\theta)|\alpha_\perp\rangle+\sin(3\theta)|\alpha\rangle,$$
$$R|\Psi_0\rangle=\cos(2\theta)|\beta_\perp\rangle+\sin(2\theta)|\beta\rangle,$$
where $|\beta\rangle\eqdef\frac{1}{\sqrt{n}}\sum_{i=1}^n|i\rangle$ and $|\beta\rangle\eqdef\frac{1}{\sqrt{(n-|x|)|x|}}\sum_{i,j:x_i=0,x_j=1}|i,j\rangle$.

Next, we investigate the special cases of $k=\kappa n, l=\lambda n$ when $1-2\kappa$ and $1-2\lambda$ are two consecutive extrema of a Chebyshev polynomial (i.e.  $\gamma-\chi=1$).
After applying Grover operators $G$ for  $d-1$ times on $|\Psi_0\rangle$, the initial state becomes
 $$ |\Psi_{d-1}\rangle\eqdef G^{d-1}|\Psi_0\rangle=\cos((2d-1)\theta)|\alpha_\perp\rangle+\sin((2d-1)\theta)|\alpha\rangle,$$
Without loss of generality, we may assume $\gamma$ is odd and continue the discussion for $\delta=0$ and $\delta=1$.
In the discussion we let $m\in\{\gamma,\gamma+1\}$ such that $\frac{m\pi}{2(2d-\delta)}=\theta=\arcsin\left(\sqrt{\frac{|x|}{n}} \right)$.
\begin{itemize}
  \item [(1)] If $\delta=1$, then $1-2\kappa=\cos\left(\frac{\gamma\pi}{2d-1}\right)$ and $1-2\lambda=\cos\left(\frac{(\gamma+1)\pi}{2d-1}\right)$  are two extrema of $T_{2d-1}(x)$. $|\Psi_{d-1}\rangle$ can be rewritten as
      $$|\Psi_{d-1}\rangle=\cos\left(\frac{m\pi}{2}\right)|\alpha_\perp\rangle+\sin\left(\frac{m\pi}{2}\right)|\alpha\rangle.$$
      Then we measure $|\Psi_{d-1}\rangle$ in computational basis, and make another query to $x_s$ if the measurement outcome is $s\in[n]$. Note that $x_s=0$ implies $m$ is not an odd integer, and $x_s=1$ implies $m$ is not an even integer. Thus we can distinguish between $|x|=k$ and $|x|=l$ with certainty.

  \item [(2)] If $\delta=0$, then $1-2\kappa=\cos\big(\frac{\gamma\pi}{2d}\big)$ and $1-2\lambda=\cos\big(\frac{(\gamma+1)\pi}{2d}\big)$  are extrema of $T_{2d}(x)$. Apply operator $R$ to $|\Psi_{d-1}\rangle$ and obtain
      $$|\Psi_{d}\rangle\eqdef R|\Psi_{d-1}\rangle=\cos(2d\theta)|\beta_\perp\rangle+\sin(2d\theta)|\beta\rangle=\cos\left(\frac{m\pi}{2}\right)|\beta_\perp\rangle+\sin\left(\frac{m\pi}{2}\right)|\beta\rangle.$$
      We now measure $|\Psi_{d}\rangle$ in orthogonal basis $\set{|k\rangle,|i,j\rangle|k,i,j\in[n],i<j}$. If the measurement result is $|k\rangle(k\in[n])$, then $m$ is not an odd integer, otherwise $m$ is not an even integer. Thus we can distinguish between $|x|=k$ and $|x|=l$ with certainty.
\end{itemize}

From the above discussion, we conclude that if $1-2\kappa$ and $1-2\lambda$ are both extrema of a Chebyshev polynomial $T_{2d-\delta}(x)$ for some $d\in[n]$ where $\delta\in\{0,1\}$ and $T_{2d-\delta}(1-2\kappa)\cdot T_{2d-\delta}(1-2\kappa)=-1$, then $f_n^{k,l}$ can be computed by a $d$-queries exact quantum algorithm.

\smallskip

One may wonder that the above argument only makes sense when both $k=\kappa n$ and $l=\lambda n$ are integers for the input length $n\in\N$.
For a general choice of parameters, the extrema of Chebyshev polynomials are likely to be irrational and hence $f_n^{k,l}$ turns out meaningless,
\textit{e.g.} for $T_4$ and $\gamma=0$, the weight decision function $f_n^{0,(1-\sqrt{2})n/2}$ is constant on its domain since no $x\in\zon$ could have an irrational weight $|x|=(1-\sqrt{2})n/2$.
However, we remark that although the above algorithm does not solve a meaningful weight decision problem, the essential power of the construction remains valid in amplifying the gap between two quantum states to be distinguished.
In \cref{general} we will introduce our ``quantum padding'' technique, which is a quantum algorithm that reduces general weight decision problems to extrema of Chebyshev polynomials as if it has the power to pad any desired number of binary digits.

\subsection{Proof of \cref{padding}}\label{general}
As discussed in \cref{relationship}, we expect to translate general $(1-2\kappa,1-2\lambda)$ (with $\kappa=\frac{k}{n},\lambda=\frac{l}{n}$ derived from $f_n^{k,l}$) to the extrema of a Chebyshev polynomial which can be computed by exact quantum algorithms.
In this subsection, we present an algorithm that can effectively pad any desired number of zeros and ones.
We first prepare a superposition  $|\Psi_0\rangle$ corresponding to the input obtained by ``padding'' $a^2$ zeros and $b^2$ ones to the input of $f_n^{k,l}$.
Namely, we introduce $a|\mathcal{L}\rangle$ and $b|\mathcal{R}\rangle$ to represent the unnormalized superpositions of newly added zeros and ones respectively.
Moreover, we define two rotations with parameters $a,b>0$  which can be used to rotate the superpositions corresponding to padded inputs into two orthogonal subspaces.
And this completes the proof of \cref{padding}.

Our algorithm will utilize two unitary transformations $W(a,b)$ and $U(a,b)$, with parameters $a,b>0$:\\
(1) $W(a,b)$ is a unitary transformation over a Hilbert space of dimension $n+2$ with basis vectors $\{|1\rangle,\ldots, |n\rangle, |\mathcal{L}\rangle, |\mathcal{R}\rangle\}$.  It is a unitary transform described as follows:
\vspace{-1mm}
      $$\left\{\begin{array}{ll}
            W(a,b)|k\rangle &= \frac{2}{n+a^2+b^2}\left(\sum_{i=1}^n|i\rangle+a|\mathcal{L}\rangle-b|\mathcal{R}\rangle\right)-|k\rangle, \forall k\in [n]; \\
            W(a,b)|\mathcal{L}\rangle &= \frac{2a}{n+a^2+b^2}\left(\sum_{i=1}^n|i\rangle+a|\mathcal{L}\rangle-b|\mathcal{R}\rangle\right)-|\mathcal{L}\rangle; \\
            W(a,b)|\mathcal{R}\rangle &= -\frac{2b}{n+a^2+b^2}\left(\sum_{i=1}^n|i\rangle+a|\mathcal{L}\rangle-b|\mathcal{R}\rangle\right)-|\mathcal{R}\rangle.
      \end{array}\right.$$
(2) $U(a,b)$ is a unitary transformation over a Hilbert space of dimension $\binom{n}{2}+3n+3$ where the basis vectors are $\{|k\rangle, |\mathcal{L}\rangle,|\mathcal{R}\rangle,|i,j\rangle, |k,\mathcal{L}\rangle,|k,\mathcal{R}\rangle,|\mathcal{L,R}\rangle|k,i,j\in[n],i<j\}$. It is a unitary completion of the following form:
\vspace{-1mm}
      $$\left\{\begin{array}{ll}
          U(a,b)|k\rangle =& \frac{1}{n+a^2+b^2}\left(\sum_{i=1}^n|i\rangle+a|\mathcal{L}\rangle+b|\mathcal{R}\rangle\right)\\
           &+\frac{1}{\sqrt{n+a^2+b^2}}\left(-\sum_{i:i<k}|i,k\rangle+\sum_{i:k<i}|k,i\rangle+a|k,\mathcal{L}\rangle+b|k,\mathcal{R}\rangle\right),k\in[n]; \\
           U(a,b)|\mathcal{L}\rangle =&\frac{a}{n+a^2+b^2}\left(\sum_{i=1}^n|i\rangle+a|\mathcal{L}\rangle+b|\mathcal{R}\rangle\right)
           +\frac{1}{\sqrt{n+a^2+b^2}}\left(-\sum_{i=1}^n|i,\mathcal{L}\rangle+b|\mathcal{L,R}\rangle\right); \\
           U(a,b)|\mathcal{R}\rangle =&\frac{b}{n+a^2+b^2}\left(\sum_{i=1}^n|i\rangle+a|\mathcal{L}\rangle+b|\mathcal{R}\rangle\right)
           +\frac{1}{\sqrt{n+a^2+b^2}}\left(-\sum_{i=1}^n|i,\mathcal{R}\rangle-a|\mathcal{L,R}\rangle\right). \\
      \end{array}\right.$$

It is easy to verify that both $W(a,b)$ and $ U(a,b)$ are unitary transformations. Similar as the Grover operator $G$ and the rotation operator $R$ in \cref{relationship} (cf. \cite{grover1996fast}), let $G(a,b)\eqdef W(a,b)O_x$ and $R(a,b)\eqdef U(a,b)O_x$. In particular, $G(a,b)$ degenerates into the standard Grover operator when $a=b=0$.
Let $a,b\geq 0$ and $ a^2=\frac{l-k}{t-s}-\frac{ls-k t}{t-s}-n$, $ b^2=\frac{ls-kt}{t-s}$. The initial state of our algorithm is $\rb{\Psi_0} $ defined as below:
$$|\Psi_0\rangle=\frac{1}{\sqrt{n+a^2+b^2}}\left(\sum_{i=1}^n|i\rangle+a|\mathcal{L}\rangle-b|\mathcal{R}\rangle\right)=\cos\theta|\alpha_\perp\rangle+\sin\theta|\alpha\rangle,$$
where $|\alpha_\perp\rangle\eqdef \frac{1}{\sqrt{n-|x|+a^2}}\left(\sum_{i:x_i=0}|i\rangle+a|\mathcal{L}\rangle\right)$, $|\alpha\rangle\eqdef \frac{1}{\sqrt{|x|+b^2}}\left(\sum_{i:x_i=1}|i\rangle-b|\mathcal{R}\rangle\right)$ and $\sin^2\theta=\frac{|x|+b^2}{n+a^2+b^2}$.
Intuitively, the above $a^2$ and $b^2$ are ``padding numbers'' of zeros and ones respectively, which can translate $k$ and $l$ into $s(n+a^2+b^2)$ and $t(n+a^2+b^2)$, even if they are not integers.
It is obvious that $a^2,b^2\geq 0$ if $(\kappa,\lambda)\in \mathsf{UL}(S_d)$.
After $d-1$ applications of $G(a,b)$ the initial state $|\Psi_0\rangle$ transforms into
$$|\Psi_{d-1}\rangle\eqdef G(a,b)^{d-1}|\Psi_0\rangle=\cos((2d-1)\theta)|\alpha_\perp\rangle+\sin((2d-1)\theta)|\alpha\rangle.$$

Without loss of generality, we assume $\gamma$ is odd.
\begin{itemize}
  \item [(1)] $s=\frac{1}{2}\left(1-\cos\left(\frac{\gamma\pi}{2d-1}\right)\right)$ and $t=\frac{1}{2}\left(1-\cos\left(\frac{(\gamma+1)\pi}{2d-1}\right)\right)$.
        Now measure the final state $|\Psi_{d-1}\rangle$ and get a measurement result $m$.
        If $m=\mathcal{L}$, return $|x|=l$; if $m=\mathcal{R}$, return $|x|=k$;
        otherwise,$m\in[n]$ and we need a query to $x_m$.
        Similarly as the discussion in \cref{relationship}, if $x_m=0$, then $|x|=l$; otherwise, $|x|=k$.

  \item [(2)] $s=\frac{1}{2}\left(1-\cos\left(\frac{\gamma\pi}{2d}\right)\right)$ and $t=\frac{1}{2}\left(1-\cos\left(\frac{(\gamma+1)\pi}{2d}\right)\right)$.
      Applying $R(a,b)$ to $|\Psi_{d-1}\rangle$ gives us
      $$|\Psi_d\rangle\eqdef R(a,b)|\Psi_{d-1}\rangle=\cos(2d\theta)|\beta_\perp\rangle+\sin(2d\theta)|\beta\rangle,$$
      where
          $$\begin{cases}
               |\beta_\perp\rangle &\eqdef\frac{1}{\sqrt{n+a^2+b^2}}\left(\sum_{i=1}^n|i\rangle+a|\mathcal{L}\rangle+b|\mathcal{R}\rangle\right),  \\
              |\beta\rangle &\eqdef\frac{1}{\sqrt{(n-|x|+a^2)(|x|+b^2)}}\left(\sum_{i:x_i=0\atop j:x_j=1}|i,j\rangle-a\sum_{i:x_i=1}|i,\mathcal{L}\rangle+b\sum_{j:x_j=0}|i,\mathcal{R}\rangle+ab|\mathcal{L,R}\rangle\right).
          \end{cases}$$

 Finally, we measure the final state $|\Psi_d\rangle$ and get a measurement result. If $m\in\{k,\mathcal{L},\mathcal{R}|k\in[n]\}$, $|x|=l$; else, $|x|=k$.
\end{itemize}
Thus we complete the proof of \cref{padding}.

%% file: exact_proof.tex

In this section, we prove the lower bounds for exact quantum query complexity of weight decision problems. Our main result is described in \cref{lowerbound} and its immediate corollary as follows.
\medskip

\noindent
\textbf{\cref{lowerbound}.}
\textit{For every $d\in\mathbb{N}$ and $0\leq k<l\leq n$ with $k,l,n\in \mathbb{N}$, let $\kappa=\frac{k}{n}$ and $\lambda=\frac{l}{n}$. If $(\kappa,\lambda)\in \mathsf{LR}(S_d)$, then $Q_E(f_n^{k,l})\geq d+1$ for sufficiently large $n$.
}


\begin{corollary}
If $n$ is sufficiently large and integers $k,l$ satisfying $0\leq k<l\leq n$, then
$Q_E(f_n^{k,l})\ge 1+\max \set{d \mid (\frac{k}{n},\frac{l}{n})\in \mathsf{LR}(S_d)}.$
\end{corollary}


In the quantum query complexity model, the degree of a Boolean function provides a lower bound for its exact quantum query complexity following Lemma \ref{beals}. Therefore, in order to prove $Q_E(f_n^{k,l})\geq d+1$, it suffices to show that $\deg(f_n^{k,l})\geq 2d+1$.
Let $p:\mathbb{R}\rightarrow\mathbb{R}$ denote the minimum degree univariate polynomial representing $f_n^{k,l}$ (see Definition~\ref{degree}), and let $q(x)= \pm(1-2p((1-x)n/2))$. 
Note that $\deg(p)=\deg(q)$,
and it suffices to prove the lower bound for $\deg(q)$.
The following two lemmas give lower bounds for
$\deg(q)$ when $0<k<l<n$ (Lemma~\ref{poly}) and $k=0$ or $l=n$ (Lemma~\ref{poly0}) respectively.

\begin{lemma}\label{poly}
For every integer $m\geq 3$ and $\gamma\in[m-2]$, let $\eta_\gamma=\cos(\frac{\gamma\pi}{m})$, $\eta_{\gamma+1}=\cos\left(\frac{(\gamma+1)\pi}{m}\right)$ and
$$
 D_\gamma \eqdef \set{(y_1,y_2)\in [-1,1]^2  \left|\begin{array}{ll}
     & (1+\eta_{\gamma+1})(y_1+1)\leq (1+\eta_{\gamma})(y_2+1)\\
     & (1-\eta_{\gamma+1})(y_1-1)\leq (1-\eta_{\gamma})(y_2-1)\\
    & y_1>y_2, (y_1,y_2)\neq (\eta_\gamma,\eta_{\gamma+1})
  \end{array}\right.
  }.
$$
For every real polynomial $p:\mathbb{R}\rightarrow \mathbb{R}$ satisfying $-1\le p(x)\le1$ for any $x \in\{1-\frac{2k}{n}|k\in[n]_0\}$, if there exists $(x_1,x_2)\in D_{\gamma}$ such that
 $p(x_1)=(-1)^{\gamma+1}$ and $p(x_2)=(-1)^\gamma$,
  then $\deg(p)\ge m+2$ when $n$ is sufficiently large.
\end{lemma}

In \cref{poly}, $\eta_\gamma$ and $\eta_{\gamma+1}$ are two consecutive extrema of the first kind Chebyshev polynomial $T_m$. $D_\gamma$ is the triangle region derived from $\mathsf{LR}\left(\frac{1}{2}(1-\eta_\gamma),\frac{1}{2}(1-\eta_{\gamma+1})\right)$ via a linear transformation.

Our proof is composed of two steps. First, we show that for any real polynomial $p$ proposed in \cref{poly}: (i) there exist two different extreme point $x'_1, x'_2$ of $p$ such that $(x'_1,x'_2)\in D_\gamma$ for sufficiently large $n$; (ii) $|p(x)|\leq 1+\eps$ for any $\eps\geq0$ and $x\in[-1,1]$ when $n$ is sufficiently large. Second, we prove that $|p(-1)|>1$ or $|p(1)|>1$ for $\deg(p)\leq m+1$ based on the first step, which is contradicted with the fact that $|p(-1)|\leq1$ and $|p(1)|\leq1$.

\begin{proof}[Proof of \cref{poly}]
For any $x\in[-1,1]$, there exists an integer $k=\lceil\frac{n-\lfloor nx\rfloor}{2}\rceil$ such that $0\le k\le n$ and $x-\frac{n-2k}{n}\le\frac{2}{n}$. Recalling that $-1\leq p((n-2k)/n)\leq 1$, by Lagrange mean value theorem, there exists $x_0\in[(n-2k)/n,x]$ such that
$$|p'(x_0)|\ge \frac{|p(x)-p((n-2k)/n)|}{x-\frac{n-2k}{n}} \ge \frac{(|p(x)|-1)n}{2}.$$
Therefore, let $|p|=\max_{|x|\le 1}|p(x)|$,
$$\max_{|x|\le 1}|p'(x)|\ge |p'(x_0)|\ge \frac{(|p|-1)n}{2}.$$
Let $d$ denote the degree of polynomial $p(x)$, and $p(x)$ has a property \cite{Achiezer} that
$$
  d^2\ge\frac{\max_{|x|\le1}|p'(x)|}{\max_{|x|\le1}|p(x)|}.
$$
That is
$$d^2\ge\frac{\max_{|x|\le1}|p'(x)|}{|p|}\ge\frac{(|p|-1)n}{2|p|},$$
and hence
 $$|p|\le1+\frac{2d^2}{n-2d^2}.$$
Namely, for any $\eps\geq0$ and sufficiently large $n$ we have
 $$|p|\le1+\varepsilon.$$

Recall that $p(x_1)=(-1)^{\gamma+1}$ and $p(x_2)=(-1)^\gamma$. Without lose of generality, let $\gamma$ be odd. If $p'(x_1)\leq 0$, there exists $x'_1\in \left[\frac{n-2\lceil n(1-x_1)/2\rceil}{n},x_1 \right]$ such that $p'(x'_1)=0$ and $p(x'_1)\geq 1$ because $p\left(\frac{n-2\lceil n(1-x_1)/2\rceil}{n}\right)\leq 1$ and $p(x_1)=1$; If $p'(x_1)\geq 0$, there exists $x'_1\in \left[x_1,\frac{n-2\lfloor n(1-x_1)/2\rfloor}{n}\right]$ such that $p'(x'_1)=0$ and $p(x'_1)\geq 1$ because $p\left(\frac{n-2\lfloor n(1-x_1)/2\rfloor}{n}\right)\leq 1$ and $p(x_1)=1$. Namely, there exists $x'_1\in (x_1-\frac{2}{n},x_1+\frac{2}{n})$ satisfying $p'(x'_1)=0$ and $p(x'_1)\geq 1$. Similarly, there exists $x'_2\in (x_2-\frac{2}{n},x_2+\frac{2}{n})$ satisfying $p'(x'_2)=0$ and $p(x'_2)\leq -1$.

Note that $(x'_1,x'_2)\in D_\gamma$ when $n$ is sufficiently large and $(x_1,x_2)\in D_\gamma$ since $D_\gamma$ is an open set. Therefore, for any $\gamma\in[m-2], $there exists $(x'_1,x'_2)\in D_\gamma$ such that
$$
\left\{
\begin{aligned}
   & (-1)^{\gamma+1}p(x_1')\ge1,\\
   & (-1)^\gamma p(x_2')\ge1,\\
   &  p'(x_1')=0,  \\
   &  p'(x_2')=0.
\end{aligned}
\right.
$$

We operate some linear transformation on polynomial $p(x)$
 and get polynomial $g(x)$, let
$$
g(x)=\frac{2(-1)^{\gamma}}{p(x_2')- p(x_1')}\left(\frac{p(x_1')+p(x_2')}{2}-p\left(\frac{(x_2'-x_1')x+x_1'\eta_{\gamma+1}-x_2'\eta_\gamma}{\eta_{\gamma+1}-\eta_{\gamma}}\right)\right).
$$
In fact, $\deg(p)=\deg(g)$ and $g(x)$ satisfies
$$
\left\{
\begin{aligned}
 & g(\eta_\gamma)=(-1)^{\gamma},\\
 & g(\eta_{\gamma+1})=(-1)^{\gamma+1},\\
 & g'(\eta_\gamma)=0,\\
 & g'(\eta_{\gamma+1})=0,\\
 & |g(x)|\le 1+\varepsilon, \forall x\in[-1,1],
\end{aligned}
\right.
$$
for any $\eps\geq 0$.

Since $(x'_1,x'_2) \in D_\gamma$, there exists $|\xi|>1$ such that $g(\xi)=\frac{2(-1)^{\gamma}}{p(x_2')- p(x_1')}\left(\frac{p(x_1')+p(x_2')}{2}-p(-1)\right)$ for $\xi=\xi_1$ or $g(\xi)=\frac{2(-1)^{\gamma}}{p(x_2')- p(x_1')}\left(\frac{p(x_1')+p(x_2')}{2}-p(1)\right)$ for $\xi=\xi_2$, where $\xi_1$ and $\xi_2$ are
   $$\xi_1=\frac{\eta_{\gamma+1}-\eta_\gamma+x'_1\eta_{\gamma+1}-x'_2\eta_\gamma}{x'_1-x'_2},$$
   $$\xi_2=\frac{\eta_\gamma-\eta_{\gamma+1}+x'_1\eta_{\gamma+1}-x'_2\eta_\gamma}{x'_1-x'_2}.$$
Next, we will prove that if $|g(\xi)|>1$, then $|p(1)|>1$ or $|p(-1)|>1$, which is a contradiction with $|p(1)|\leq1$ or $|p(-1)|\leq1$. Without loss of generality, we assume that $\gamma$ is odd. When $|g(\xi)|>1$ with $\xi=\xi_1$, we have $p(-1)>p(x'_1)\ge 1$ or $p(-1) < p(x'_2)\le -1$, i.e., $|p(-1)|>1$.  When $|g(\xi)|>1$ with $\xi=\xi_2$, we have $p(1)>p(x'_1)\ge 1$ or $p(1) < p(x'_2)\le -1$, i.e., $|p(1)|>1$.

In the rest of this proof, we will show that if $\deg(p)\leq m+1$, then $|g(\xi)|>1$.

Let $T_m(x)$ denote the $m$-th Chebyshev polynomial of the first kind. Because $\deg(g)=\deg(p)\leq m+1$, let $h(x)\eqdef g(x)-T_m(x)=\sum_{i=0}^{m+1}a^*_ix^i$, where $a^*_i\in \mathbb{R}$ for all $i\in\{0,\ldots,m+1\}$. According to Property \ref{chebyproperty}, we have
\begin{equation}\label{linearconstraints1}
\left\{
  \begin{array}{ll}
     & h(\eta_\gamma)=0, \\
     & h(\eta_{\gamma+1})=0, \\
     & h'(\eta_\gamma)=0, \\
     & h'(\eta_{\gamma+1})=0, \\
     & (-1)^kh\left(\cos \left(\frac{k\pi}{m}\right)\right)\le\varepsilon, 0\le k\le m.
  \end{array}
\right.
\end{equation}
Let ${\bf{a}}=[a_0,\ldots,a_{m+1}]^\mathrm{T} \in \mathbb{R}^{m+2}$. For any $i\in\{0,\ldots,m+1\}$, let ${\bf{e}_i}=[e_{i0},\ldots,e_{i,m+1}]^\mathrm{T}\in\mathbb{R}^{m+2}$ where $e_{ii}=1$ if $a^*_i\ge 0$, $e_{ii}=-1$ if $a^*_i < 0$ and $e_{ij}=0$ for all $i\neq j$. Let $\boldsymbol{\eps}=[\eps_0,\ldots,\eps_{m+8}]^\mathrm{T}$ in which $\eps_i=\eps$ for $0\leq i \leq m$ and $\eps_j=0$ for $m+1\leq j \leq m+8$. Let ${\bf C}\in \mathbb{R}^{(m+9)\times (m+2)}$ where the row index set and column index set are $\{0,\ldots,m+8\}$ and $\{0,\ldots,m+1\}$ respectively. The matrix ${\bf C}$ can be defined as follows:

$${\bf C}=\left\{
    \begin{array}{ll}
      {\bf C}_{ij}=(-1)^i\cos^j(\frac{i\pi}{m}), & \mbox{if~} 0\leq i\leq m, 0\leq j \leq m+1;\\
      {\bf C}_{m+1+k,j}=(-1)^{\gamma+k}\cos^j\left(\frac{(\gamma+k)\pi}{m}\right), & \mbox{if~} 0\leq j \leq m+1, k\in\{0,1\};\\
      {\bf C}_{m+3,0}={\bf C}_{m+4,0}=0, & \\
      {\bf C}_{m+3+k,j}=j\cos^{j-1}\left(\frac{(\gamma+k)\pi}{m}\right), & \mbox{if~} 1\leq j \leq m+1,k\in\{0,1\}; \\
      {\bf C}_{m+k+4,j}=-{\bf C}_{m+k,j}, & \mbox{if~} 1\leq k \leq 4, 0\leq j\leq m+1.
    \end{array}
  \right.
$$
Define a linear programming:
$$
\begin{array}{ll}
  \max~~~~ & {\bf{e}_i}^\mathrm{T} {\bf a}; \\
   s.t.~~~~ & {\bf Ca}\leq \boldsymbol{\eps};
\end{array}
$$
where $|a_i|={\bf{e}_i}^\mathrm{T} {\bf a}$ for any $0\leq i\leq m+1$. Since ${\bf{a}}^*=[a^*_0,\ldots,a^*_{m+1}]^\mathrm{T}$ satisfies linear constraints (\ref{linearconstraints1}), ${\bf{a}}^*$ is a feasible solution of linear programming. Let ${\bf b}=[b_0,\ldots,b_{m+1}]^\mathrm{T}$. Then, the corresponding asymmetric dual problem is
$$
\begin{array}{ll}
  \min~~~~ & \boldsymbol{\eps}^\mathrm{T} {\bf b}; \\
   s.t.~~~~ & {\bf C}^\mathrm{T}{\bf b}={\bf{e}_i}; \\
           & {\bf b}\geq 0.
\end{array}
$$
If $\eps=0$, then $h(x)$ has at least $m+2$ roots based on linear constraints (\ref{linearconstraints1}). Since $\deg(h)\leq m+1$, $h(x)\equiv 0$. Namely, $[0,\ldots,0]^\mathrm{T}\in \mathbb{R}^{m+2}$ is a feasible solution of primal problem. Hence, there exists a feasible solution ${\bf b}^*$ of dual problem where ${\bf C}^\mathrm{T}{\bf b}^*={\bf{e}_i}$ and ${\bf b}^*$ is independent with $\boldsymbol{\eps}$.
Recall that ${\bf{a}}^*$ and ${\bf{b}}^*$ are feasible solutions of primal and dual problem, respectively.
If $\eps>0$, then $|a^*_i|={\bf{e}_i}^\mathrm{T} {\bf a}^* \leq \boldsymbol{\eps}^\mathrm{T} {\bf b}^*$ according to weak duality theorem \cite{bot2009duality}. Therefore, $|a^*_i|\leq \boldsymbol{\eps}^\mathrm{T} {\bf b}^* \leq \eps\|{\bf b}^*\|_1 $ for any $0\leq i\leq m+1$. Then $|h(\xi)|=|\sum_{i=0}^{m+1}a^*_i\xi^i|\leq \sum_{i=0}^{m+1}|a^*_i||\xi|^i\leq \eps \|{\bf b}^*\|_1 \sum_{i=0}^{m+1}|\xi|^i$. If $|\xi|>1$, $|T_n(\xi)|>1$ due to Property \ref{chebyproperty}. Let $\eps=\frac{|T_m(\xi)|-1}{2\|{\bf b}^*\|_1 \sum_{i=0}^{m+1}|\xi|^i}>0$, then $|g(\xi)|=|h(\xi)+T_m(\xi)|\geq |T_m(\xi)|-|h(\xi)|>(|T_m(\xi)|+1)/2>1$.\qed

\end{proof}

\begin{lemma}\label{poly0}
For every polynomial $p:\mathbb{R}\rightarrow \mathbb{R}$ satisfying $-1\le p(x)\le1$ for all $x \in\{1-\frac{2k}{n}|k\in[n]_0\}$.
If there exists an $x_0$ satisfying either of the follows conditions:
\begin{enumerate}
  \item  $x_0\in(\cos(\frac{\pi}{m}),1)$ such that $p(x_0)=-1$ and $p(1)=1$;
  \item  $x_0\in(-1,-\cos(\frac{\pi}{m}))$ such that $p(x_0)=1$ and $p(-1)=-1$;
\end{enumerate}
then $\deg(p)\geq m+1$ when $n$ is sufficiently large.
\end{lemma}

The proof of Lemma \ref{poly0} is similar to Lemma \ref{poly}, and is given in \cref{poly0proof}.
The proof of \cref{lowerbound} combines Lemma \ref{poly}, Lemma \ref{poly0}, together with Lemma~\ref{beals}.

\begin{proof}[Proof of \cref{lowerbound}]					
	Let polynomial $p:\mathbb{R}\rightarrow \mathbb{R}$ denote the minimum degree univariate polynomial representation of $f_n^{k,l}$.
	Then $p$ satisfies $p(k)=0$, $p(l)=1$ and $0\leq p(i)\leq 1$ for all $i\in[n]$. For every $d\in \mathbb{N}$ and $(\kappa,\lambda)\in \mathsf{LR}(S_d)$, there exists $(s,t)\in S_d$ satisfying $(\kappa,\lambda)\in \mathsf{LR}(s,t)$. Suppose $s=\frac{1}{2}\left(1-\cos\left(\frac{\gamma \pi}{2d-\delta}\right)\right)$ and $t=\frac{1}{2}\left(1-\cos\left(\frac{(\gamma+1) \pi}{2d-\delta}\right)\right)$  where $\gamma\in[2d-1]_0$ if $\delta=0$ and $\gamma\in[2d-3]$ if $\delta=1$.
	Let $q(x)=(-1)^{\gamma+1}(1-2p((1-x)n/2))$, where $\deg(p)=\deg(q)$. It follows that $q(1-2k/n)=(-1)^{\gamma+1}$, $q(1-2l/n)=(-1)^\gamma$ and $-1\leq q(1-2i/n)\leq 1$ for all $i\in[n]_0$.
	If $\delta\in\{0,1\}$ and $\gamma\in[2d-\delta-1]$, we have $Q_E(f_n^{k,l})\geq d+1$ following Lemma  \ref{poly} and Lemma \ref{beals};
	otherwise if $\delta=0$ and $\gamma\in\{0,2d-1\}$, we have $Q_E(f_n^{k,l})\geq d+1$ following Lemma \ref{poly0} and Lemma \ref{beals}.\qed
\end{proof}					


\begin{remark}
There are two probable reasons why our upper and lower bounds are not perfectly matched: i) \cref{lowerbound} only makes use of the degree of Chebyshev polynomials (indeed stretched Chebyshev polynomials after padding), whereas the best known lower bound Lemma \ref{beals} is not restricted to Chebyshev polynomials; ii) the lower bound by Lemma \ref{beals} may not be tight, given the recent result that degree-$4$ polynomials are not equivalent to $2$-query quantum algorithms in the bounded-error model \cite{arunachalam2017quantum}.
In what follows we elaborate on the reasons with three sets of $(\frac{k}{n},\frac{l}{n})$ estimating the boundary of $Q_E(f_n^{k,l})=d$ derived from \cref{padding,lowerbound} and Lemma \ref{beals} respectively (for simplicity we assume both $k$ and $l$ are integers):\\
$$\left\{
\begin{array}{ll}
	\mathcal{A}_d&\!\!\!\!\!:=\mathsf{UL}(S_d);\\
 	\mathcal{B}_d&\!\!\!\!\!:=\mathsf{LR}(S_d);\\
 	\mathcal{C}_d&\!\!\!\!\!:=\set{(\kappa,\lambda)\in I^2\;|\; \exists \text{ polynomial $p:[0,1]\rightarrow[0,1]$, s.t.} \deg(f)\leq 2d,f(\kappa)=0,f(\lambda)=1}.
\end{array}\right.
$$

Specifically, $Q_E(f_n^{k,l})\le d$ for every  $(\frac{k}{n},\frac{l}{n})\in\mathcal{A}_d$ by \cref{padding},
while $Q_E(f_{n}^{k,l}) > d$ for $(\frac{k}{n},\frac{l}{n})\in \mathcal{B}_d$ by \cref{lowerbound}
and $Q_E(f_n^{k,l})>d$ if $(\frac{k}{n},\frac{l}{n}) \in \overline{\mathcal{C}_d}$ by Lemma \ref{beals} ($\overline{\mathcal{B}_d}$ and $ \overline{\mathcal{C}_d}$ are the complementaries of $\mathcal{B}_d$ and $\mathcal{C}_d$).
In fact there is $\mathcal{A}_d\subseteq \mathcal{C}_d \subsetneq \overline{\mathcal{B}_d}$ for $d>1$ and furthermore it is likely $\mathcal{A}_d\subsetneq \mathcal{C}_d$.
As a result $\mathcal{A}_d\union \mathcal{B}_d$ does not cover all the choices of $(\kappa,\lambda)$, and hence we cannot determine from \cref{padding,lowerbound} whether $Q_E(f_n^{k,l})\le d$ or $Q_E(f_n^{k,l})> d$ as long as $(\frac{k}{n},\frac{l}{n})\in \overline{ \mathcal{A}_d\union \mathcal{B}_d}$,
even if it is the case $(\frac{k}{n},\frac{l}{n})\in \overline{\mathcal{C}_d}\intersect {\overline{\mathcal{B}_d}} = \overline{\mathcal{C}_d} \intersect (\overline{\mathcal{A}_d\union \mathcal{B}_d}) $ when  $Q_E(f_n^{k,l})> d$ by Lemma \ref{beals}.
%
%
Anyhow, upper and lower bounds for exact quantum query complexity of $f_n^{k,l}$ can still be derived from \cref{padding,lowerbound} but with a larger gap, and fortunately the gap is no more than one in most cases as verified numerically, \emph{i.e.} $\union_{d\in\N} \left( \overline{\mathsf{A}_{d+1} \union \mathcal{B}_d}\right)$ is small.
\end{remark}

%% file: append.tex
\section{Exact quantum algorithm for $f_n^{k,l}$}\label{alg}
The exact quantum algorithm for $f_n^{k,l}$ is described in \cref{alg:WD}.
\begin{algorithm}[!htb]\normalsize
\caption{Weight decision algorithm for $f_n^{k,l}$}\label{alg:WD}
\begin{algorithmic}[1]
\REQUIRE $x=(x_1,\ldots,x_n)\in \zon$.
\ENSURE $f_n^{k,l}(x)$. \\
$k,l$ satisfies $(\frac{k}{n},\frac{l}{n})\in \mathsf{UL}(s,t)$, where $(s,t)\in S_d$.\\
Suppose $(s,t)=\left(\frac{1}{2}\left(1-\cos\left(\frac{\gamma \pi}{2d-\delta}\right)\right),\frac{1}{2}\left(1-\cos\left(\frac{(\gamma+1) \pi}{2d-\delta}\right)\right)\right)$, where $\delta\in\{0,1\}$ and $\gamma\in \{0,\ldots,2d-\delta\}$.\\
\STATE \textbf{initialization:} Prepare a superposition
$|\Psi_0\rangle\eqdef \frac{1}{\sqrt{n+a^2+b^2}}\left(\sum_{i=1}^n|i\rangle+a|\mathcal{L}\rangle-b|\mathcal{R}\rangle\right)$, \\ where  $a^2=\frac{l-k}{t-s}-\frac{ls-k t}{t-s}-n$ and $b^2=\frac{ls-kt}{t-s}$.
\STATE Apply $G(a,b)$ $d-1$ times to $|\Psi_0\rangle$:
$|\Psi_{d-1}\rangle\eqdef G^{d-1}(a,b)|\Psi_0\rangle.$
\IF{$\gamma$ is odd}
\IF {$\delta=1$}
\STATE Measure $|\Psi_{d-1}\rangle$ in basis $\{|i\rangle,|\mathcal{L}\rangle,|\mathcal{R}\rangle|i\in[n]\}$ and obtain a measurement result $m$.
\STATE \textbf {if} $m$ is $\mathcal{L}$, \textbf{return} 1;\\
       \textbf {if} $m$ is $\mathcal{R}$, \textbf{return} 0;\\
       \textbf {if} $m\in \{k|k\in [n]\}$, query $x_k$ and \textbf{return} $1-x_k$.\\
\ELSE
\STATE Apply $R(a,b)$ to $|\Psi_{d-1}\rangle$:
$|\Psi_d\rangle\eqdef R(a,b)|\Psi_{d-1}\rangle$.
\STATE Measure $|\Psi_{d}\rangle$ in basis $\{|k\rangle,|k,\mathcal{L}\rangle,|k,\mathcal{R}\rangle,|i,j\rangle,|\mathcal{L}\rangle,|\mathcal{R}\rangle,|\mathcal{L,R}\rangle|k,i,j\in[n],i<j\}$ and obtain a measurement result $m$.
\STATE \textbf{if}  $m\in\{k,\mathcal{L},\mathcal{R}|k\in[n]\}$, \textbf{return} 1.\\
\textbf{if}  $m\in\{(i,j),(k,\mathcal{L}),(k,\mathcal{R}),(\mathcal{L,R})|k,i,j\in[n],i<j\}$, \textbf{return} 0.
\ENDIF
\ELSE
\IF {$\delta=0$}
\STATE Measure $|\Psi_{d-1}\rangle$ in basis $\{|i\rangle,|\mathcal{L}\rangle,|\mathcal{R}\rangle|i\in[n]\}$ and obtain a measurement result $m$.
\STATE \textbf {if} $m$ is $\mathcal{L}$, \textbf{return} 0;~~
       \textbf {if} $m$ is $\mathcal{R}$, \textbf{return} 1;~~
       \textbf {if} $m\in \{k|k\in [n]\}$, query $x_k$ and \textbf{return} $x_k$.\\
\ELSE
\STATE Apply $R(a,b)$ to $|\Psi_{d-1}\rangle$:
$|\Psi_d\rangle\eqdef R(a,b)|\Psi_{d-1}\rangle$.
\STATE Measure $|\Psi_{d}\rangle$ in basis $\{|k\rangle,|k,\mathcal{L}\rangle,|k,\mathcal{R}\rangle,|i,j\rangle,|\mathcal{L}\rangle,|\mathcal{R}\rangle,|\mathcal{L,R}\rangle|k,i,j\in[n],i<j\}$ and obtain a measurement result $m$.
\STATE \textbf{if}  $m\in\{k,\mathcal{L},\mathcal{R}|k\in[n]\}$, \textbf{return} 0;\\
\textbf{if}  $m\in\{(i,j),(k,\mathcal{L}),(k,\mathcal{R}),(\mathcal{L,R})|k,i,j\in[n],i<j\}$, \textbf{return} 1.
\ENDIF
\ENDIF

\end{algorithmic}
\end{algorithm}

\section{Exact quantum query complexity of a special function}\label{specialfunction}
It may be observed in the proof of \cref{lowerbound} that for $0\le k<\frac{n}{2}$, the minimum-degree polynomial computing $f_n^{k,\frac{n}{2}}$ also computes a symmetric partial Boolean function $g_n^k$ defined as follows:
$$g_n^k(x)=\left\{
  \begin{array}{ll}
    0, & |x|\in\set{k,n-k}; \\
    1, & |x|=\frac{n}{2}; \\
    undefined, & \hbox{otherwise.}
  \end{array}
\right.$$
Furthermore, we note that for $k=0$, $g_n^{0}(x)$ is exactly the Deutsch-Jozsa function \cite{deutsch1992rapid} and hence $Q_E(g_n^{0})=1$;
whereas for $d>1$ and $\frac{1}{2}\left(1-\cos\left(\frac{(d-2)\pi}{2(d-1)}\right)\right)<\frac{k}{n}\leq \frac{1}{2}\left(1-\cos\left(\frac{(d-1)\pi}{2d}\right)\right)$,
$Q_E(g_n^k)=d$ if $n$ is sufficiently large.
The proof is the same as \cref{padding} and \cref{lowerbound}.

\section{Proof of Lemma \ref{poly0}}\label{poly0proof}
The proof sketch is the same with Lemma \ref{poly}.
Let $d$ denote the degree of $p(x)$.
Similar with the proof of \cref{poly}, when $n$ is sufficiently large, we have
$$
\max_{|x|\le1}|p(x)|\le1+\varepsilon, \forall \eps>0.
$$
Without generality, we only discuss case (1). Recall that $p(x_0)=-1$ for $x_0\in (\cos(\frac{\pi}{m}),1)$ and $p(1)=1$. If $p'(x_0)\leq 0$, there exists $x'_0\in\left[x_0,\frac{n-2\lfloor(1-x_0)n/2\rfloor}{n}\right]$ satisfying $p'(x'_0)=0$ and $p(x'_0)\leq -1$; If $p'(x_0)\geq 0$, there exists $x'_0\in\left[\frac{n-2\lceil(1-x_0)n/2\rceil}{n},x_0\right]$ satisfying $p'(x'_0)=0$ and $p(x'_0)\leq -1$. Namely, there exists $x'_0\in (x_0-\frac{2}{n},x_0+\frac{2}{n})$ satisfying $p'(x'_0)=0$ and $p(x'_0)\leq -1$. If $n$ is sufficiently large, there exists $x'_0\in (\cos(\frac{\pi}{m}),1)$ satisfying $p(x'_0)\leq -1$, $p'(x'_0)=0$ since $x_0\in (\cos(\frac{\pi}{m}),1)$.
Define a polynomial $g(x)$ by $p(x)$:
$$
g(x)=\frac{2}{1-p(x_0')}\left(\frac{1+p(x_0')}{2}-p\left(\frac{(x_0'-1)x+\cos (\frac{\pi}{m})-x_0'}{\cos (\frac{\pi}{m})-1}\right)\right).
$$
In fact, $\deg(p)=\deg(g)$. Moreover, $g(1)=1$, $g(\cos (\frac{\pi}{m}))=-1$, $g'(\cos (\frac{\pi}{m}))=0$ and $-1-\varepsilon\le g(x)\le 1+\varepsilon$ for all $x\in[-1,1]$ and $\varepsilon>0$.
If $x'_0\in(\cos(\frac{\pi}{m}),1)$, there exists $\xi=\frac{2\cos(\frac{\pi}{m})-x'_0-1}{1-x'_0}<-1$ such that $g(\xi)=\frac{2}{1-p(x_0')}\left(\frac{p(x_0')+1}{2}-p(-1)\right)$.
If $|g(\xi)|>1$, then $p(-1)\leq p(x'_0)<-1$ or $p(-1)>1$.  In the rest of this proof, we will show that if $p(x)$ is a polynomial of degree at most $m$, then $|g(\xi)|>1$. Namely, $|p(-1)|>1$ holds when $\deg(p)\leq m$, which is a contradiction with $|p(-1)|\leq1$. Therefore, $\deg(p)\geq m+1$.

Let $h(x)=g(x)-T_m(x)=\sum_{i=0}^{m}a^*_ix^i$, where $T_m(x)$ is the $m$-th Chebyshev polynomial of the first kind. Based on Lemma \ref{poly}, we have
\begin{equation}\label{linearconstraints2}
\left\{
\begin{aligned}
&h(1)=0,\\
&h\left(\cos\left(\frac{\pi}{m}\right)\right)=0,\\
&h'\left(\cos\left(\frac{\pi}{m}\right)\right)=0,\\
&(-1)^kh\left(\cos\left(\frac{k\pi}{m}\right)\right)\le\varepsilon, 0\le k\le m.
\end{aligned}
\right.
\end{equation}

Let ${\bf{a}}=[a_0,\ldots,a_{m}]^\mathrm{T} \in \mathbb{R}^{m+1}$. For any $i\in\{0,\ldots,m\}$, let ${\bf{e}_i}=[e_{i0},\ldots,e_{i,m}]^\mathrm{T}\in\mathbb{R}^{m+1}$ where $e_{ii}=1$ if $a^*_i\ge 0$, $e_{ii}=-1$ if $a^*_i < 0$ and $e_{ij}=0$ for all $i\neq j$. Let $\boldsymbol{\eps}=[\eps_0,\ldots,\eps_{m+6}]^\mathrm{T}$ in which $\eps_i=\eps$ for $0\leq i \leq m$ and $\eps_j=0$ for $m+1\leq j \leq m+6$. Let ${\bf C}\in \mathbb{R}^{(m+7)\times (m+1)}$ where the row index set and column index set are $\{0,\ldots,m+6\}$ and $\{0,\ldots,m\}$ respectively. The matrix ${\bf C}$ can be defined as follows:

$${\bf C}=\left\{
    \begin{array}{ll}
      {\bf C}_{ij}=(-1)^i\cos^j\left(\frac{i\pi}{m}\right), & \mbox{if~} 0\leq i\leq m, 0\leq j \leq m;\\
      {\bf C}_{m+1+k,j}=(-1)^k\cos^j\left(\frac{k\pi}{m}\right), & \mbox{if~} 0\leq j \leq m, k\in\{0,1\};\\
      {\bf C}_{m+3,0}=0, & \\
      {\bf C}_{m+3,j}=j\cos^{j-1}\left(\frac{\pi}{m}\right), & \mbox{if~} 1\leq j \leq m; \\
      {\bf C}_{m+k+3,j}=-{\bf C}_{m+k,j}, & \mbox{if~} 1\leq k \leq 3, 0\leq j\leq m.
    \end{array}
  \right.
$$

Define a linear programming:
$$
\begin{array}{ll}
  \max~~~~ & {\bf{e}_i}^\mathrm{T} {\bf a}; \\
   s.t.~~~~ & {\bf Ca}\leq \boldsymbol{\eps};
\end{array}
$$
where $|a_i|={\bf{e}_i}^\mathrm{T} {\bf a}$ for any $0\leq i\leq m+1$. Since ${\bf{a}}^*=[a^*_0,\ldots,a^*_{m}]^\mathrm{T}$ satisfies linear constraints (\ref{linearconstraints2}), ${\bf{a}}^*$ is a feasible solution of linear programming. Let ${\bf b}=[b_0,\ldots,b_{m}]^\mathrm{T}$. Then, the corresponding asymmetric dual problem is
$$
\begin{array}{ll}
  \min~~~~ & \boldsymbol{\eps}^\mathrm{T} {\bf b}; \\
   s.t.~~~~ & {\bf C}^\mathrm{T}{\bf b}={\bf{e}_i}; \\
           & {\bf b}\geq 0.
\end{array}
$$
Similar discussion with Lemma \ref{poly}, there exists a feasible solution ${\bf b}^*$ of dual problem which is independent with $\boldsymbol{\eps}$. According to weak duality theorem, $|a^*_i|\leq \boldsymbol{\eps}^\mathrm{T} {\bf b}^* \leq \eps\|{\bf b}^*\|_1 $ for any $0\leq i\leq m$. Therefore, $|h(\xi)|=|\sum_{i=0}^{m}a_i\xi^i|\leq \sum_{i=0}^{m}|a_i||\xi|^i\leq \eps \|{\bf b}^*\|_1 \sum_{i=0}^{m}|\xi|^i$. Because $|T_m(\xi)|>1$ for $\xi<-1$ duo to Property \ref{chebyproperty}, let $\eps=\frac{|T_m(\xi)|-1}{2\|{\bf b}^*\|_1 \sum_{i=0}^{m}|\xi|^i}>0$, then $|g(\xi)|=|h(\xi)+T_m(\xi)|\geq |T_m(\xi)|-|h(\xi)|>(|T_m(\xi)|+1)/2>1$.